\documentclass[twocolumn,aps,pra,showpacs]{revtex4}

\usepackage{epsfig}
\usepackage[latin1]{inputenc}
\usepackage[T1]{fontenc}
\usepackage[english]{babel}
\usepackage{graphicx}
\usepackage{amssymb}
\usepackage{amsmath}
\usepackage{amscd}
\usepackage{eucal}
\usepackage{color}
\usepackage{bm}
\usepackage{amsfonts}

\newcommand{\li}{{\mathrm{L}}}
\newcommand{\di}{{\mathrm{Da}}}
\newcommand{\la}{\langle}
\newcommand{\ra}{\rangle}
\newcommand{\br}{{\bf r}}
\newcommand{\bp}{{\bf p}}

\begin{document}

\title{Long-Time Coherence in Echo Spectroscopy with
  $\pi/2$--$\pi$--$\pi/2$ Pulse Sequence}

\author{Arseni Goussev} \affiliation{School of Mathematics, University
  of Bristol, University Walk, Bristol BS8 1TW, UK}

\author{Philippe Jacquod} \affiliation{Physics Department, University
  of Arizona, 1118 E. 4$^{\rm th}$ Street, Tucson, AZ 85721, USA}

\begin{abstract}
  Motivated by atom optics experiments, we investigate a new class of
  fidelity functions describing the reconstruction of quantum states
  by time-reversal operations as $ M_\di(t) = | \la \psi | e^{i H_2 t
    / 2} e^{i H_1 t / 2} e^{-i H_2 t / 2} e^{-i H_1 t / 2} | \psi \ra
  |^2$.  We show that the decay of $M_\di$ is quartic in time at short
  times, and that it freezes well above the ergodic value at long
  times, when $H_2-H_1$ is not too large. The long-time saturation
  value of $M_\di$ contains easily extractable information on the
  strength of decoherence in these systems.
\end{abstract}

\maketitle


\section{Introduction}

When subjected to external noisy fields, quantum mechanical
wavefunctions lose memory of their phase. As a fundamentally important
consequence of this {\it decoherence} process, pairs of partially
scattered waves no longer interfere, and the dynamics follows the
Liouville time-evolution of classical densities~\cite{Joos03}.  A
somehow similar situation occurs when one time-evolves an initial
superposition $\phi = \sum_\alpha c_\alpha \psi_\alpha$ of many
eigenmodes $\psi_\alpha$ of the Hamiltonian $H_1$ governing the time
evolution, with incommensurate eigenfrequencies $\epsilon_\alpha$. In
this case, for each pair of components $(\alpha,\beta)$, the relative
phase $(\epsilon_\alpha-\epsilon_\beta) t$ becomes pseudo-random,
which washes out partial wave interferences.  This {\it dephasing}
process, however, differs from decoherence in a fundamental way that
it can in principle be undone by an appropriate time-inversion.  As a
matter of fact, echo experiments are able to reverse the sign of the
Hamiltonian, $H_1 \rightarrow -H_1$, by means of effective changes of
coordinate axes induced by electromagnetic pulses~\cite{hahn}.  When
this operation is performed after an evolution time $t$, one expects
the initial wavefunction to be reconstructed at $2t$, regardless of
its spread over eigenmodes.  Imperfections in the pulse sequence or
unavoidable couplings to external uncontrolled degrees of freedom
result instead in an imperfect time-inversion, $H_1 \rightarrow
-H_2=-H_1-\Sigma$, and therefore the Loschmidt
echo~\cite{peres84,jalabert01,cerruti,gorin06,jacquod09} (we set
$\hbar \equiv 1$)
\begin{subequations}
\label{eq:01}
\begin{align}
  M_\li(t) &= |m_\li(t)|^2 \, , \quad \mathrm{with}\label{eq:01.a} \\
m_\li(t)& =\langle \psi |
e^{i H_2 t} e^{-i H_1 t}|  \psi \rangle \, ,\label{eq:01.b}
\end{align}
\end{subequations}
gives a better description of the fidelity with which the experiment
reconstructs the initial state. Echo experiments in nuclear magnetic
resonance~\cite{hahn,nmr}, quantum optics~\cite{kurnit},
atomic~\cite{buch,davidson,davidson2}, condensed
matter~\cite{nakamura}, microwave cavities~\cite{schaefer}, and
elastodynamics~\cite{weaver} have demonstrated that $M_{\rm L}(t)$
remains sizable for times significantly longer than the dephasing
time. The decay of $M_{\rm L}(t)$ allows one to extract information on
irreversible decoherence processes induced by $\Sigma$.

In experiments with cold atoms the Loschmidt echo, $M_\li$, can be
extracted from interference fringes of Ramsey
spectroscopy~\cite{comment}. There, an effectively two-level atom is
initially prepared in a state $| 1 \rangle \otimes | \psi \rangle$,
where $| 1 \rangle$ and $| 2 \rangle$ denote the two internal atomic
states, and $| \psi \rangle$ stands for the spatial component of the
initial state. First, the atom is irradiated with a microwave
frequency field with energy chosen to change the atomic state into an
equiprobable superposition of $| 1 \rangle \otimes | \psi \rangle$ and
$| 2 \rangle \otimes | \psi \rangle$. Such a field is referred to as a
$\pi/2$ pulse. The atom is then let to evolve in an optical trap for a
time $t$ during which the $| 1 \rangle$-component of the state evolves
under a spatial Hamiltonian $H_1$, while the $| 2 \rangle$-component
under $H_2$. After that another $\pi/2$ pulse is applied to the atom
and the probability $P_2$ for the atom to be found in the internal
state $| 2 \rangle$ is measured. It turns out that this probability is
essentially determined by the Loschmidt echo amplitude, $m_\li$. In
practice however one works not with a pure initial state $| \psi
\rangle$, but with a thermal mixture of initial states. The echo
amplitude $m_\li$ from each of these states contributes to $P_2$ with
a different, effectively random phase, which in turn reduces the
fringe contrast in a Ramsey experiment. As a result, the
$\pi/2$--$\pi/2$ pulse sequence proves inefficient in measuring the
Loschmidt echo for large ensembles of thermally populated states.

In order to overcome this difficulty Davidson and collaborators
implemented a novel pulse sequence in their echo spectroscopy
experiments \cite{davidson,davidson2}
\begin{subequations}
\label{eq:02}
\begin{align}
  M_\di(t) &= |m_\di(t)|^2 \, , \quad \mathrm{with} \label{eq:02.a}\\
  m_\di(t) &= \la \psi | e^{i H_2 t / 2} e^{i H_1 t / 2} e^{-i H_2 t /
    2} e^{-i H_1 t / 2} | \psi \ra \, . \label{eq:02.b}
\end{align}
\end{subequations}
The corresponding pulse sequence consisted of three short pulses,
$\pi/2$--$\pi$--$\pi/2$, separated by two time intervals of equal
duration $t/2$, after which $P_2$ was measured. The $\pi$ pulse swaps
the population of the internal states $| 1 \rangle$ and $| 2
\rangle$. The probability $P_2$ is then determined by the amplitude
$m_\di$, and each individual state of the thermal ensemble contributes
to $P_2$ with the same phase. Thus, the $\pi/2$--$\pi$--$\pi/2$ pulse
sequence allows one to measure the echo in Eq.~\eqref{eq:02} even for
ensembles of more than $10^6$ of thermally populated states, as in the
experiments of Refs.~\cite{davidson,davidson2}.

It is clear from the definitions given by Eqs.~\eqref{eq:01} and
\eqref{eq:02} that mathematically $M_\di$ is not the same quantity as
the Loschmidt echo $M_\li$. Even though some significant differences
between $M_\di$ and $M_\li$ have been previously envisaged in the
literature they have never been systematically studied. It is the
purpose of this article to fill in this gap by comparing the two
quantities both analytically and numerically. Below, we show that
$M_\di$ differs from the Loschmidt echo $M_\li$ in the two important
respects that (i) its short-time decay is quartic and not quadratic in
time, and (ii) for not too strong perturbation $\Sigma = H_2-H_1$, $
M_\di$ saturates at a perturbation-dependent value, well above the
ergodic saturation of $ M_\li(\infty) \sim N^{-1}$ at the inverse
Hilbert space size. Fidelity freezes have been reported for Loschmidt
echoes with off-diagonal perturbations with a zero time
average~\cite{prosen}, phase-space displacement
perturbations~\cite{cyril}, and more recently for initially pure
states coupled to complex environments~\cite{kohler}, however the
freeze we report here has a different physical origin. We note in
particular that it persists for $t \rightarrow \infty$. The long-time
saturation of $M_\di$ allows to extract the strength of the fields in
$\Sigma$ more easily than by fitting decay curves of conventional
echoes, over not precisely defined time intervals. Moreover, the
absence of decay arising from $\Sigma$ in this pulse sequence makes it
straightforward to extract decoherence rates because, assuming that
the pulse sequence is perfect, any decay in experimentally obtained
data for $M_{\rm Da}(t)$ would come exclusively from the coupling of
the system to external degrees of freedom, not included in our
theory. Given the superb experimental control that modern echo
experiments have on their pulse sequence, this novel echo spectroscopy
has therefore the potential to deliver precious, previously
unattainable information on the dominant sources of decoherence in
trapped cold atomic gases.



\section{Short-time decay}

There has been a large number of analytical and numerical
investigations of the Loschmidt echo and some of its
offsprings~\cite{gorin06,jacquod09}. Most, if not all approaches
assume a small perturbation, i.e. $|\Sigma| \ll |H_{1,2}|$ for an
appropriate operator norm. As but one consequence, the largest energy
scale is the energy bandwidth $B$, which to leading order is the same
for $H_1$ and $H_2$.  For short times, $t \ll B^{-1}$, $M_\li(t)$ is
easily calculated by expanding the propagators in Eqs.~\eqref{eq:01},
and keeping the leading order contributions. One obtains
\begin{equation}
\label{eq:03}
  M_\li(t) \simeq 1 -(\sigma_\li t)^2 \,,
\end{equation}
where
\begin{equation}
\label{eq:04}
  \sigma_\li^2 = \la \psi | \Sigma_\li^2 | \psi \ra
  - \la \psi | \Sigma_\li | \psi \ra^2 \,, \quad \Sigma_\li = H_1-H_2 \,.
\end{equation}
Thus, the short-time decay of the Loschmidt echo is {\it quadratic}
\cite{peres84,cerruti}, with a rate given by the dispersion
$\sigma_\li$ of the perturbation operator $\Sigma_\li$ evaluated over
the initial state.

The same procedure can be applied to $M_\di$, where it however gives
\begin{equation}
  M_\di(t) \simeq 1 -(\sigma_\di t)^4 \,,
\label{eq:05}
\end{equation}
with the decay rate $\sigma_\di$ given by
\begin{equation}
  \sigma_\di^4 = \la \psi | \Sigma_\di^2 | \psi \ra
  - \la \psi | \Sigma_\di | \psi \ra^2 \,, \quad 
  \Sigma_\di = \frac{i}{4} [H_1,H_2] \,.
\label{eq:06}
\end{equation}
Two things are remarkable here. First, the short-time decay of $M_\di$
is {\it quartic} in $t$, and thus slower than the decay of
$M_\li$. Second, its rate is determined by the
{\it commutator} of the unperturbed and perturbed Hamiltonians.


\section{Long-time saturation}

The analysis of the long-time behavior of $M_\li$ and $M_\di$ starts
by diagonalizing the unperturbed and perturbed Hamiltonian operators,
$H_1 = \sum_u E_u | u \ra \la u |$ and $H_2 = \sum_v E_v | v \ra \la v
|$, and expanding the initial state in the basis of the unperturbed
Hamiltonian, $| \psi \ra = \sum_u c_u | u \ra$~\cite{jacquod09}.  The
resulting expression for the echo is then averaged over time to yield
the mean saturation value.  In the case of the Loschmidt echo the
time-averaged saturation is given by
\begin{equation}
  M_{\li,\infty} = \!\!\! \sum_{u,u',u'',v} c_u^* c_{u''} |c_{u'}|^2
  \la u | v \ra \la v | u' \ra \la u' | v \ra \la v | u'' \ra \, .
\label{eq:08}
\end{equation}
The next step is to average this expression over a random ensemble of
coefficient $c_u$ for the initial state, such that $\overline{c_u^*
  c_{u'}} = N^{-1} \delta_{u,u'}$. (Hereinafter, an overline denotes
the averaging over an ensemble of random initial states.) Here $N$ is
the effective size of the Hilbert space (the number of eigenstates of
$H_{1,2}$ comprising the initial state).  To leading order in $1/N$,
one uses $\overline{c_u^* c_{u''} |c_{u'}|^2} = \overline{c_u^*
  c_{u''}} \cdot \overline{|c_{u'}|^2} = N^{-2} \delta_{u,u''}$ to
obtain the ergodic saturation value
\begin{equation}
  \overline{M_{\li,\infty}} = \frac{1}{N} \,.
\label{eq:08.5}
\end{equation}
Using the same procedure, one can calculate the long-time saturation value
of $M_\di$. At the level
of the echo amplitude $m_\di$, one gets
\begin{subequations}
\label{eq:09}
\begin{align}
  \overline{m_{\di,\infty}} &= \sum_{u,u',v} \overline{c_u^* c_{u'}}
  \la u | v \ra \la v | u' \ra \la u' | v \ra \la v | u' \ra \label{eq:09.a}\\
  &= \frac{1}{N} \sum_{u,v} \big| \la u | v \ra \big|^4
  \,.\label{eq:09.b}
\end{align}
\end{subequations}
One then uses an approximation $\overline{\big| \la u | v \ra \big|^4}
\simeq \overline{\big| \la u | v \ra \big|^2}^2$ with $
\overline{\big| \la u | v \ra \big|^2}= \rho(E_u-E_v)$ a function of
only the energy difference between the two states. Replacing one of
the sums in Eq.~\eqref{eq:09} by an integral over the energy
difference between the two states scaled by the mean level spacing
$\Delta=B/N$, we can write
\begin{equation}
\overline{m_{\di,\infty}} 
  \simeq \int \, \frac{{\rm d} E}{\Delta} \rho^2(E)\, .
\label{eq:09.c}
\end{equation}
This expression relates the long-time saturation of $M_\di$ to the 
energy spreading of 
eigenfunctions of $H_1$ over those of $H_2$ as measured by $\rho(E)$. 
It is known for a large variety of quantum chaotic systems that, in the regime 
$\Delta \ll \Gamma /\Delta  \ll B$, this spreading has a 
Lorentzian shape
\begin{equation}
\label{eq:10}
 \rho(E_u-E_v)\simeq \frac{\Delta}{\pi}
  \frac{\Gamma/2}{(E_u-E_v)^2 + (\Gamma/2)^2} \,,
\end{equation}
with a spreading width given by the golden rule, $\Gamma \simeq
\sigma_{\rm L}^2/\Delta$ \cite{gruver97}, see Eq.~(\ref{eq:04}) for the
definition of $\sigma_{\rm L}^2$.  We thus obtain
\begin{equation}
\overline{m_{\di,\infty}} 
  \simeq \frac{\Delta}{\pi \Gamma}\,.
\label{eq:11}
\end{equation}
Eqs.~\eqref{eq:10} and \eqref{eq:11} predict an average saturation
value $\overline{M_{\di,\infty}}$ above the ergodic saturation for
$N<(B/\pi \Gamma)^2$. The width $\Gamma$ of the Lorentzian
(\ref{eq:10}) increases with $|\Sigma|$, and the ergodic saturation,
Eq.(\ref{eq:08.5}), is recovered when $\Gamma > B/\pi N^{1/2}$, thus
\begin{eqnarray}\label{eq:sat}
  \overline{M_{\di,\infty}} \simeq 
  {\rm max} \left[\left( \frac{\Delta}{\pi \Gamma} \right)^2, \frac{1}{N} \right] \, .
\end{eqnarray}
We note that the Lorentzian spreading of Eq.~(\ref{eq:10})
is replaced by more complicated, system-dependent spiked structures in dynamical systems with
mixed or regular dynamics, for which it is accordingly impossible to draw general conclusions.
We stress, however, that Eq.~(\ref{eq:09.c}) remains valid even in that case.

Eq.~(\ref{eq:sat}) is the main result of this
paper. This new long-time fidelity saturation originates from
the specific sequence of time-evolutions
in $M_\di$, giving the long-time behavior of the latter
as an energy integral over the 
{\it squared} average overlap $|\langle u|v \rangle|^4$
of eigenstates $|u\rangle$ of $H_1$
over the eigenstates $|v\rangle$ of $H_2$.
For completeness we next comment on the intermediate regime,
between the short-time quartic decay and the long-time saturation.


\section{Intermediate asymptotic decay}

We briefly sketch a semiclassical analysis of $M_\di$ in the
intermediate regime between the short-time decay and the long-time
saturation.  We follow the lines of Ref.~\cite{jalabert01} to show
that $M_\di$ and $M_\li$ have the same behavior in that regime.

\begin{figure}[t]
\centerline{\epsfig{figure=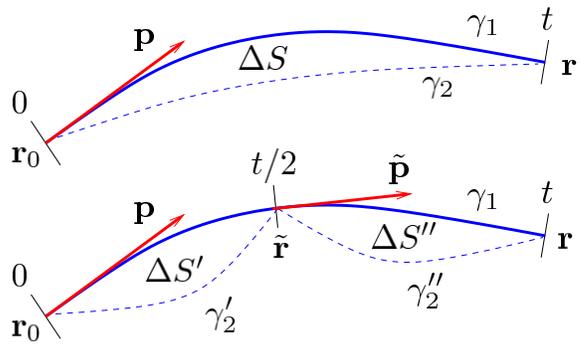,width=3in}}
\caption{(Color online) Trajectories of the unperturbed ($\gamma_1$)
  and perturbed ($\gamma_2$, $\gamma'_2$, and $\gamma''_2$) systems
  together with the associated action differences ($\Delta S$, $\Delta
  S'$, and $\Delta S''$). }
\label{fig1}
\end{figure}

In the semiclassical approximation the time evolution of 
$|\psi\ra$ under $H_j$, $j=1,2$, is given by
\begin{equation}
  \la \br | e^{-i H_j t} | \psi \ra = 
  \! \int d\br' \!\!\!\! \sum_{\gamma(\br' \rightarrow \br, t)}
  \!\!\! D_{j,\gamma} \, e^{i S_{j,\gamma}} \la \br' | \psi \ra \,.
\label{eq.van_vleck}
\end{equation}
Here, the sum goes over all classical paths $\gamma$ connecting $\br'$
and $\br$ in time $t$, $S_{j,\gamma} = S_{j,\gamma}(\br,\br',t)$ is
the action along $\gamma$, $D_{j,\gamma} = (2\pi i)^{-d/2}
|\mathrm{det}(\partial^2 S_{j,\gamma} / \partial \br \partial
\br')|^{1/2} e^{-i \pi \nu_{j,\gamma}/2}$ with Morse index
$\nu_{j,\gamma}$ counting the number of conjugate points on $\gamma$,
and $d$ is the dimensionality of the system \cite{gutzwiller}. The
semiclassical Loschmidt echo amplitude is obtained by inserting
Eq.~\eqref{eq.van_vleck} in Eq.~\eqref{eq:01.b}. The resulting
expression contains three spatial integrals over $\br$, $\br'$, and
$\br''$ with a double sum over trajectories $\gamma_1(\br' \rightarrow
\br, t)$ and $\gamma_2(\br'' \rightarrow \br, t)$ corresponding to the
Hamiltonians $H_1$ and $H_2$ respectively. The standard analysis of
this expression involves three steps~\cite{jalabert01}: (i) One
assumes that $\la \br | \psi \ra$ is localized about a point $\br_0$
and evaluates the integrals over $\br'$ and $\br''$ by stationary
phase approximations. This reduces the set of paths $\gamma_1$ and
$\gamma_2$ to those starting at $\br_0$, see Fig.~\ref{fig1}. (ii)
Noting that the double sum over trajectories contains rapidly
oscillating phase factors $\exp[i(S_{1,\gamma_1}-S_{2,\gamma_2})]$, so
that only pairs of correlated paths $\gamma_1$ and $\gamma_2$
contribute to $m_\di$, one employs the diagonal approximation
($\gamma_2 \simeq \gamma_1$) to reduce $m_\li$ to a sum over a single
path $\gamma_1$. Ref.~\cite{vanicek}, building up on ideas first
expressed in Ref.~\cite{cerruti}, justified this step by the shadowing
theorem.  (iii) Finally, one uses the fact that $|D_{1,\gamma_1}|^2$
is the Jacobian of a transformation between final positions $\br$ and
initial momenta $\bp$ on paths $\gamma_1$. This allows one to change
the integration variable from $\br$ to $\bp$ to get
\begin{equation}
  m_\li(t) = (2\pi)^{-d} \int d\bp \; e^{i \Delta S} \: \big| \la \bp | \psi \ra \big|^2 \,.
\label{eq:06.5}
\end{equation}
Here $\Delta S = \Delta S(\br_0, \bp, t) =
S_{1,\gamma_1}-S_{2,\gamma_2}$ is the difference between the action of
an unperturbed trajectory $\gamma_1$ leaving the point $\br_0$ with a
momentum $\bp$ and traveling for time $t$ and the action of the
corresponding perturbed trajectory $\gamma_2 \simeq \gamma_1$.
Following the same procedure one finds
\begin{equation}
  m_\di(t) = (2\pi)^{-d} \int d\bp \; e^{i (\Delta S'-\Delta S'')} \: \big| \la \bp | \psi \ra \big|^2 \,,
\label{eq:06.6}
\end{equation}
where $\Delta S' = \Delta S(\br_0,\bp, t/2)$ and $\Delta S'' = \Delta
S(\tilde{\br},\tilde{\bp}, t/2)$ with $(\tilde{\br},\tilde{\bp})$
being the phase space point on $\gamma_1$ at time $t/2$, see
Fig.~\ref{fig1}. In other words, $\Delta S'$ ($\Delta S''$) is the
action difference between the first (second) half of the unperturbed
trajectory $\gamma_1$ and the corresponding perturbed trajectory
$\gamma'_2$ ($\gamma''_2$). This is sketched in Fig.~\ref{fig1}.

Once averaged over an ensemble of initial states, both $M_{\li}$ and
$M_{\di}$ satisfy
\begin{align}
  \overline{M_{\li,\di}(t)} &\simeq \overline{m_{\li,\di}(t)}^{\, 2} \label{eq:06.7}\\
  &+ (2\pi)^{-2d} \int d\bp \int_{\Omega_\bp} \!\! d\bp' \,
  \big| \la \bp | \psi \ra \big|^2 \, \big| \la \bp' | \psi \ra \big|^2 \,, \nonumber 
\end{align}
where the integral over ${\bf p}'$ is restricted to a volume
$\Omega_{\bp}$ around ${\bf p}$, such that two trajectories starting
from the same spatial point with momenta $\bp$ and $\bp' \in
\Omega_{\bp}$ stay ``close'' in phase space during time $t$. The first
term in the right-hand side of Eq.~\eqref{eq:06.7} is evaluated using
the central limit theorem, $\overline{\exp(i\Delta S)} \simeq
\exp(-\overline{\Delta S^2}/2) \simeq e^{-\Gamma t/2}$ and
$\overline{\exp[i(\Delta S'-\Delta S'')]} \simeq
\exp[-(\overline{\Delta S'^2}+\overline{\Delta S''^2})/2] \simeq
e^{-\Gamma (t/2+t/2)/2} = e^{-\Gamma t/2}$, where $\Gamma$ is defined
in Eq.~(\ref{eq:10}) as the width of the local density of states.  For
$M_\di$, we neglect correlations between $\Delta S'$ and $\Delta S''$,
which is justified by the fast decay of correlations along chaotic
classical trajectories.  The second term in Eq.~\eqref{eq:06.7} is
determined by the measure of the set $\Omega_\bp$ and in chaotic
systems decays as $e^{-\lambda t}$ with $\lambda$ being the average
Lyapunov exponent of the underlying classical system
\cite{jalabert01}. Therefore, the intermediate time decay of $M_{\di}$
is the same as that of $M_{\li}$ \cite{jalabert01,jacquod01}, i.e.
\begin{equation}
  \overline{M_\li(t)} \simeq \overline{M_\di(t)} 
  \sim e^{-t \min[\Gamma,\lambda]} \,.
\label{eq:07}
\end{equation}
This exponential time decay continues until the echo reaches the
saturation plateau given by Eq.~\eqref{eq:sat}.

\begin{figure}[t]
\centerline{\epsfig{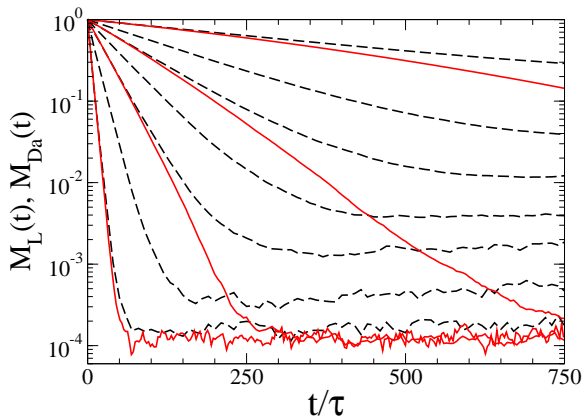}}
\caption{(Color online) Average echo decay for the kicked rotator
  model with $K_1=57$, $N=8192$, and $K_2-K_1=5 \cdot 10^{-5}$, $1.2
  \cdot 10^{-4}$, $2.1 \cdot 10^{-4}$ and $5 \cdot 10^{-4}$ ($M_\li$,
  red solid lines from top to bottom), and $K_2-K_1=5 \cdot 10^{-5}$,
  $9 \cdot 10^{-5}$, $1.2 \cdot 10^{-4}$, $1.6 \cdot 10^{-4}$, $2.1
  \cdot 10^{-4}$, $3.1 \cdot 10^{-4}$, and $5 \cdot 10^{-4}$ ($M_\di$,
  black dashed lines, from top to bottom). Curves are averages over
  500 initial states.}
\label{fig2}
\end{figure}

\bigskip

Ref.~\cite{davidson} reported some saturation of $M_\di$ for
ultra-cold atoms inside optical traps. However, at this stage, a
direct comparison of these experiments with our theory does not seem
feasible, because they explore completely different time
regimes. Indeed, the echo spectroscopy experiments of
Refs.~\cite{davidson,davidson2} are concerned with short times
corresponding to no more than 3-4 oscillations/bounces of an atom in
the trap.  In contrast, the semiclassical derivation of the
exponential decay, Eq.~\eqref{eq:07}, and the RMT analysis of the
fidelity freeze, Eq.~\eqref{eq:sat}, are only valid for times much
longer than the average free flight time.


\section{Numerical study}

We confirm our analytical results with some numerical data. Our
simulations are based on the kicked rotator model with dimensionless
Hamiltonian
\begin{equation}\label{kickrot}
H_{1,2} = \frac{\hat{p}^2}{2} + K_{1,2} \cos \hat{x} \sum_n \delta(t-n \tau).
\end{equation}
For large enough 
kicking strength, $K_{1,2} \tau > 7$, the dynamics is
fully chaotic with a Lyapunov exponent $\lambda = \ln[K_{1,2} \tau /2]$.  We
quantize this Hamiltonian on a torus, and accordingly consider
discrete values $p_l=2 \pi l/N$ and $x_l=2 \pi l/N$, $l=1,...N$,
giving an effective Planck's constant $\hbar_{\rm eff}= 1/N$. Both
echoes $M_\li(n)$ and $M_\di(n)$ are computed for discrete times $t=n
\tau$, with the kicking period $\tau$, using the unitary Floquet
operators 
$U_{1,2}=\exp[-i \hat{p}^2/2 \hbar_{\rm eff}] \exp[-iK_{1,2}
\cos \hat{x}/ \hbar_{\rm eff}]$ for single-kick time-evolutions.  
The bandwidth is $B=2 \pi$ and accordingly $\Delta = 2 \pi/N$.
The eigenstates of $U_2$
spread over those of $U_1$ according to Eq.~(\ref{eq:10}) with $\Gamma
\propto (\delta K \, N)^2$, with $\delta K = K_2-K_1$~\cite{jacquod09}.  
Together with
Eq.~(\ref{eq:sat}), we thus expect a long-time saturation of $M_\di$
at a value
\begin{eqnarray}
\label{satur_prediction}
\overline{M_{\di,\infty}} \sim \left(\delta K^2 N^3  \right)^{-2} \, ,
\end{eqnarray}
for $\delta K^4 N^5 < 1$.

\begin{figure}[t]
\centerline{\epsfig{figure=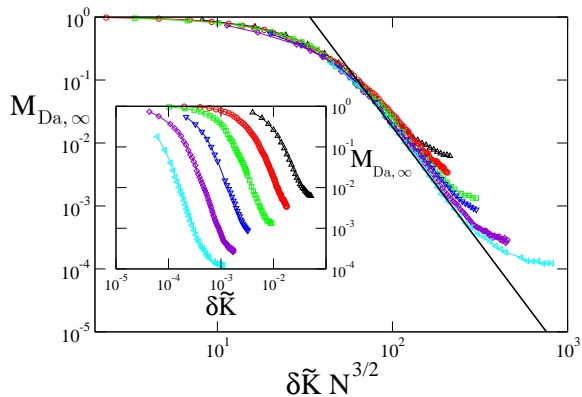,width=3in}}
\caption{(Color online) Long-time saturation value of $M_{\di}$ for
  $K_1=57$ and $N=256$ (black up triangles), 512 (red circles), 1024
  (green squares), 2048 (blue down triangles), 4096 (violet diamonds)
  and 8192 (cyan left triangles).  Main panel: rescaled data
  confirming the analytical prediction of
  Eq.(\ref{satur_prediction}). The straight black line indicates a
  slope of $\propto 1/x^{3.8}$. Inset: raw data as function of the
  difference of dimensionless kicking strengths $\delta
  \tilde{K}=\tilde{K}_2-\tilde{K}_1$, with
  $\tilde{K}_{1,2}=K_{1,2}/\tau$.}
\label{fig3}
\end{figure}

Fig.~\ref{fig2} shows the time decay of the echoes,
$\overline{M_\li(t)}$ shown as red curves and $\overline{M_\di(t)}$ as
black curves, averaged over an ensemble of randomly chosen initial
states. For equal values of the perturbation strength, both $M_\li$
and $M_\di$ display an exponential time decay governed by the same
decay rate, providing a clear support for Eq.~\eqref{eq:07}. The
Loschmidt echo decay saturates at a value $\sim N^{-1}$ in agreement
with Eq.~\eqref{eq:08.5}. The freeze of $M_{\di}$ occurs at a value
that decreases with increasing perturbation strength until it reaches
ergodic saturation at $N^{-1}$.  We confirm in Fig.~\ref{fig3} that
the numerically observed perturbation-dependent saturation of $M_\di$
follows Eq.~\eqref{satur_prediction}. Once plotted as a function of
$\delta K N^{3/2}$, saturation data for $N \in [256,8192]$ and $\delta
K \in [4 \cdot 10^{-5},0.052]$ nicely fall on top of one another until
they deviate because they have different ergodic saturation,
$N^{-1}$. Moreover, in the regime of validity $\Delta \ll \delta K \ll
B$ of Eq.~(\ref{eq:10}), one has $M_{\di,\infty} \propto (\delta K
N^{3/2})^b$ with an exponent $b \simeq 3.8$ close to the prediction
$b=4$ from Eq.~(\ref{satur_prediction}). We note that $b$ is larger
for data with larger Hilbert space size $N$, where the fitting range
is larger -- and the fit is accordingly more accurate -- because
saturation occurs at larger values of $\delta K N^{3/2}$. We also
checked numerically that the initial decay of $M_\di$ is quartic and
not quadratic in time. Our numerical simulations thus fully confirm
the theoretical predictions derived above.

\section{Conclusions}

Our analysis of the new fidelity function
$M_\di$~\cite{davidson,davidson2} shows that it significantly differs
from the Loschmidt echo in two important respects: (i) the short-time
decay of $M_\di$ is quartic (and not quadratic) in time, and is
governed by the commutator (and not the difference) of the unperturbed
and perturbed Hamiltonians, and (ii) for not too strong Hamiltonian
perturbations, the decay of $M_\di$ freezes at values inversely
proportional to the square of the measure $\Gamma$ of the
perturbation, as defined by the width of the local density of states,
Eq.~(\ref{eq:10}). This allows to estimate the strength of decoherence
processes in systems of cold trapped atoms by fitting the saturation
value of $M_\di$, which is arguably easier and more precise than
fitting decay curves over not precisely defined time intervals.
In addition to providing an analytic derivation of this finding, in
particular relating the saturation level to the strength of
decoherence fields, and to predicting an initial quartic decay of
$M_\di$, our theory gives an intermediate behavior of $M_\di$ which
follows that of the Loschmidt echo $M_\li$. We confirmed these
analytical findings numerically.


\emph{Acknowledgments.--} A.G. acknowledges the support by EPSRC under
Grant No. EP/E024629/1.


\end{document}